# Zero-index Weyl metamaterials

Farzad Zangeneh-Nejad and Romain Fleury*

*Laboratory of Wave Engineering, Swiss Federal Institute of Technology in Lausanne (EPFL), 1015 Lausanne, Switzerland*

*To whom correspondence should be addressed. Email: romain.fleury@epfl.ch

**Depending on the geometry of their Fermi surfaces, Weyl semimetals and their analogues in classical systems have been classified into two types. In type I Weyl semimetals (WSMs), the cone-like spectrum at the Weyl point (WP) is not tilted, leading to a point-like closed Fermi surface. In type II WSMs, on the contrary, the energy spectrum around the WP is strongly tilted such that the Fermi surface transforms from a point into an open surface. Here, we demonstrate, both theoretically and experimentally, a new type of (classical) Weyl semimetal whose Fermi surface is neither a point nor a surface, but a flat line. The distinctive Fermi surfaces of such semimetals, dubbed as type III or zero-index WSMs, gives rise to unique physical properties: one of the edge modes of the semimetal exhibits a zero index of refraction along a specific direction, in stark contrast to types I and II WSMs for which the index of refraction is always non-zero. We show that the zero-index response of such topological phases enables exciting applications such as extraordinary wave transmission.**

Weyl semimetals (WSMs) have recently attracted enormous research interest because of their unconventional band structures and topological features [1-4]. In the Brillouin zone of a WSM, the valence and conduction bands touch each other at a set of points, called Weyl points (WPs) [5-11]. These points are sources and drains of Berry curvature in the momentum space and, thus, can be associated with a topological charge [12]. The topological nature of WPs protects them against annihilation in case of small perturbations, and leads to a large variety of intriguing physical phenomena such as superconductivity [13,14], chiral anomaly [14,16], and topological negative refraction [17].

Although it is not possible to remove Weyl nodes because of their topological character, the energy dispersion at a WP might, in principle, be tilted along a certain direction. While such a tilting of the band structure does not affect the essential topology of the WP, it can change the bulk properties of the semi-metallic phase by modifying the geometry of the Fermi surface [18]. As such, depending on whether their cone-like band structure is tilted or not, WSMs and their classical analogues have been grouped into two types. In a standard (type I) WSM [19,20], the energy dispersion at the WP is not tilted, leading to a point-like Fermi surface. In type II WSMs [21-28], on the other hand, the energy spectrum around the Weyl node is strongly tilted (by an angle larger than a specific angle $\theta_c$, at which parts of the conduction and valence bands start to coexist in energy). As a result, the cone-like dispersion around the WP tips over and the Fermi surface transforms from a single point into an open surface. This gives rise to distinctive physical properties for type II WSMs such as orientation dependent chiral anomaly and anomalous Hall effects [29,30].

So far, Weyl semimetals have been studied both when the tilting angle at the WP is smaller than $\theta_c$ (type I WSMs), or larger (type II WSMs). Here, on the contrary, we investigate in theory, simulation and experiment, the exotic properties of classical Weyl semimetals when their energy spectrum is tilted exactly by $\theta_c$. We show that the Fermi surface of such kinds of semi-metallic phases is neither a point, as in type I WSMs, nor a surface, as in type II WSMs. Instead, it is a line connecting a pair of WPs. Since the geometry of the Fermi surface plays a very important role in determining the bulk properties of WSMs, we categorize these new kinds of semimetals into a distinct group, referred to as type III or "zero-index" WSMs. Such an appellation is inspired by the fact that, as opposed to type I and II WSMs possessing a non-zero index of refraction in all directions, the topological surface states of the proposed WSM exhibit a zero group velocity, or more precisely, a zero-index response along a specific direction. This behavior generalizes to three-dimensions the zero-index response of the edge modes of type III Dirac semimetals [23]. We demonstrate that the static-like behavior of type III Weyl semimetals can be used to achieve extra-ordinary wave transmission, akin to anomalous tunneling in the so-called zero-index metamaterials [31-34].

To start, we consider the tight-binding crystal shown in Fig. 1a, described by the tight-binding Hamiltonian

$$H = 2\lambda_z \cos k_z \sigma_0 + (2\lambda_x \cos k_x + 2\lambda_y \cos k_y)\sigma_x + 2\delta_x \sin k_x \sigma_y + 2\delta_z \cos k_z \sigma_z \quad (1)$$

in which $\sigma_0$, $\sigma_{x,y,z}$ are identity and Pauli matrices, respectively, and $k_x$, $k_y$, and $k_z$ are the Bloch wave numbers. We assume $\lambda_x = 1, \lambda_y = 2, \lambda_z = 2$, and $\delta_x = 0.5$. Furthermore, for now, we suppose that the parameter $\delta_z$ is set to be zero, i.e. $\delta_z = 0$. With these choices of parameters, the tight-binding Hamiltonian $H$ vanishes at $P: (k_x, k_y, k_z) = (0, \pm 2\pi/3, \pm\pi/2)$, implying that the dispersion bands of the two-level system cross each other at these four points. As an example, we report in Fig. 1b the energy band structure of the corresponding Hamiltonian at the plane $k_y = 2\pi/3$, which includes two of these crossing points at $(k_x, k_z) = (0, \pm\pi/2)$. These crossing points are Weyl nodes, possessing opposite chiralities (see [35]). The energy dispersion around these Weyl nodes is not tilted, leading to a point-like closed Fermi surface (characteristic of type I WSMs).

Next, suppose that the parameter $\delta_z$ is set to be $\delta_z = 2\lambda_z$. The corresponding band structure at the plane $k_y = 2\pi/3$ is shown in Fig. 1c. Like the previous case, the band structure exhibits a pair of Weyl points located at $(k_x, k_z) = (0, \pm\pi/2)$. However, as opposed to the previous case, the cone-spectrum is significantly tilted around these nodes, so that some modes from the valence band coexist in energy with others in the conduction band. Such a strong tilting modifies the geometry of the Fermi surface: it becomes an open surface (characteristic of type II WSMs [21]).

Now we assume $\delta_z = \lambda_z$. Fig. 1d represents the corresponding band structure at $k_y = 2\pi/3$. Similar to the previous two cases, the band structure exhibits two Weyl nodes at $(k_x, k_z) = (0, \pm\pi/2)$. Around these points, the energy band structure is tilted, similar to what we had in the latter case. However, the tilting angle is exactly equal to the critical angle $\theta_c$, leading to a Fermi surface which is neither a single point, like type I WSMs, nor an open surface, like type II WSMs, but a straight line connecting a pair of Weyl points [35]. Such a unique Fermi surface gives rise to unique physical properties for the semi-metallic phase (referred to as type III WSM), as demonstrated below.

We first investigate the topological boundary states carried by zero-index Weyl semimetals. To this end, we consider a 100×1×1 supercell of the crystal and calculate the corresponding dispersion surfaces [35]. The resulting band structure is represented in Fig. 2a, clearly showing the existence of two helical gap-closing topological edge states. In order to further analyze this result, we report the dispersion curves corresponding to three different plane sections, namely $k_y = \pi$, $k_y = 2\pi/3$ and $k_y = 0$ (Figs. 2b-d, respectively). As explained earlier, the plane $k_y = 2\pi/3$ is the one at which the Weyl transition occurs, so the band structure is gapless at this plane (Fig. 2c). As confirmed by direct calculations, the constant-$k_y$ planes before ($k_y = 0$) and after ($k_y = \pi$) this critical point possess different $\mathbb{Z}_2$ topological invariants, making the system a $\mathbb{Z}_2$ Weyl semimetal [35]. In particular, $k_y = \pi$ has a zero invariant [35], consistent with the fully gapped dispersion diagram of Fig. 2b. Conversely, $k_y = 0$ is associated with a non-vanishing invariant [35], explaining the two helical topological surface states observed in Fig. 2d. Remarkably, in contrast to type I and II WSMs for which the gap-less states have always a non-zero group velocity, one of the associated helical topological surface states of the obtained WSM has a zero group velocity along $k_z$ (note that in directions other than $k_z$, the edge mode remains dispersive). The origin of this behavior is linked to the balance between the $\sigma_0$ and $\sigma_z$ terms in Eq. 1, canceling one of the diagonal elements of the Hamiltonian at the critical angle.

Next, we investigate an implementation of the proposed zero-index WSM using the sonic crystal shown in Fig. 3a, consisting of Helmholtz resonators connected to each other via acoustic channels with varying widths (see [35] for a geometrical description). Full-wave simulations [35] confirm that the band structure of the crystal displays a type III Weyl semi-metallic phase (Fig. 3b). Fig. 3c represents the band structure of a $11 \times 1 \times 1$ super-cell of the crystal. Consistent with our prior findings, the band structure of the super-cell consists of plane sections with different topologies. At the topological plane ($k_y = k_{y3}$), the crystal supports two helical surface states, one of which has a near-zero index of refraction along $z$ (note that, due to longer range hopping, the associated effective index is not exactly zero, but near it). To obtain more insights into the physics of the near-zero-index topological edge mode, we plot in Fig. 3d (left panel) the associated acoustic pressure field, obtained via eigen-frequency simulations [35]. It is seen that, along the $z$ direction, this mode has a quasi-static pressure distribution. This is equivalent to an infinite phase velocity along the edge of the semimetal. Interestingly, such a zero-index behavior also occurs at another cut-plane of the band structure, namely the transition plane $k_y = k_{y2}$. This is demonstrated in Fig. 3d (right), reporting the associated acoustic field profile at this plane for $k_z = 0$.

The static-like behavior of type III (zero-index) topological phases enables exciting physical phenomena. Fig. 3e, for example, demonstrates the possibility of achieving extra-ordinary transmission of sound [36] by direct anomalous matching of an external waveguide to the near-zero topological edge mode of the $k_y = 0$ 2D topological cut-plane of the Weyl semimetal. Due to the small value of both the effective density and bulk modulus of the edge channel, the edge mode is impedance-matched to the external waveguides (when neglecting losses, see [35]), leading to the transmission amplitude of near unity (0.988) with almost no phase lag (0.052 rad). Note that, markedly different from Fabry-Pérot resonances, the obtained anomalous resonance tunneling does not depend on the length of the semi-metallic connection [34,35].

Now, we experimentally investigate such anomalous tunneling through the edge mode at a topological cut-plane, based on the fabricated prototype of the sonic crystal shown in Fig. 4a-b. Note that, as explained in [35], picking this particular 2D cut-plane allows building only one-layer of the 3D system. This one-layer system may also be viewed as a type III Dirac semimetal [23], supporting a 1D zero-index topological edge mode. Two external waveguides are connected to the edge of the fabricated prototype, with a loudspeaker at the entrance of the first and an anechoic termination in the second. Four microphones probe the acoustic pressure distribution along the edge of the semimetal. Full wave simulations (neglecting losses) predict zero-index tunneling of sound through the edge at the frequency $f_0 = 2.92\ kHz$. In practice, due to the presence of absorption, the transmission coefficient of the fabricated structure does not reach unity at this frequency. However, the existence of the edge mode at $f_0$ and its static-like phase profile are not affected by the losses, and can be probed in experiment [35]. Fig. 4c (top) reports the sound pressure level spectrum measured by microphone number 3. Near $f_0$, a resonance peak is observed, corresponding to the near-zero topological edge mode. The quasi-static nature of the edge mode is confirmed in Fig. 4c (bottom), reporting the phases of the pressure field at all four microphones versus frequency. As observed, near $f_0$, the corresponding phase spectra converge to almost the same value. To further illustrate the near-zero-index response of the topological edge mode, we report in Fig. 4d the variation of the pressure phase at $f_0$ along the edge of the crystal (red region), together with the standard phase variation in waveguides, i.e. $\varphi = 2\pi f_0 L/c$ (green region). This figure confirms the quasi-static sound propagation along the edge of the semimetal with almost no phase variations along a distance of 1.4 acoustic wavelengths.

Finally, leveraging the concept of synthetic dimensions [37-40], we measure the complete band structure of the type III Weyl semimetal. To this end, we consider a one-dimensional sonic crystal made from evanescently coupled acoustic bound states in continuum (BIC) [41-45] with on-site frequencies $\omega_n$ and hopping rates $k_n$ following the equations: $\omega_n = 2\lambda_z \cos\varphi_z (1 + (-1)^n)$ and $k_n = \lambda_x + (-1)^n \delta_x + \lambda_y \cos\varphi_y$ [35]. It is straightforward to verify that, upon replacing $\varphi_y$ with $k_y$ and $\varphi_z$ with $k_z$, the tight-binding Hamiltonian of this one-dimensional chain is identical to the one of Eq. 1. Therefore, the complete semi-metallic band structure can be probed considering the family of one-dimensional sonic systems generated when sweeping the parameters $\varphi_y$ and $\varphi_z$ in a synthetic 3D Brillouin zone. We report in Figs. 5a-c the band structures of the chains versus $k_z(\varphi_z)$ at three different $k_y(\varphi_y)$, namely $k_y = \pi$, $k_y = 2\pi/3$, and $k_y = 0$, respectively, obtained by full-wave eigen-frequency simulations. Consistent with our previous tight-binding studies, at $k_y = \pi$, the energy spectrum is gapped. Contrarily, at $k_y = 2\pi/3$, it is gap-less and exhibits a pair of Weyl points, connected to each other with a flat line. At $k_y = 0$, which is a plane with a non-zero $\mathbb{Z}_2$ charge, the two gap-closing helical surface states are indeed found, one of them with a zero group velocity.

To extract the band structure of the crystal from far-field scattering tests, we excite the system under study with a plane-wave and measure the corresponding transmission spectra by performing standard standing-wave pattern analysis [35]. The evolution of the transmission coefficient of the structure as a function of both frequency and $k_z$ is plotted in Figs. 5d-f, corresponding respectively to $k_y = 0$ (panel d), $k_y = 2\pi/3$ (panel e), and $k_y = \pi$ (panel f). It is observed that the distinctive characteristics of the band structure of the crystal, including the peculiar dispersion of the topological edge modes (Fig. 5a-c), translate into clear features in the scattering spectrum (Fig. 5d-f), namely transmission maxima or minima. These findings are verified in experiment, based on one-dimensional sonic crystals of plastic rods arranged inside an acoustic waveguide [35]. By changing the radii of the rods and the distance between them, the on-site frequencies and hopping rates of the 1D BIC chain were tuned according to the prescribed relations, so as to resolve the dispersion bands of the proposed semimetal. The results of our experiment, shown in Fig. 5g-I, agree well with simulation.

To conclude, we explored the properties of zero-index WSMs and demonstrated their intriguing dispersion properties in theory, simulation, and experiment. The unconventional character of the topological boundary states of zero-index WSM may open exciting venues for a large variety of engineering-oriented applications, for example energy concentration, supercoupling, wavefront shaping and subwavelength imaging. Future studies may investigate non-Hermitian effects that could arise in the strong absorption regime, such as the creation of exceptional rings and their effect on the edge states. This may be experimentally reached by designing geometries supporting larger viscothermal losses [48-50].

**Figures**

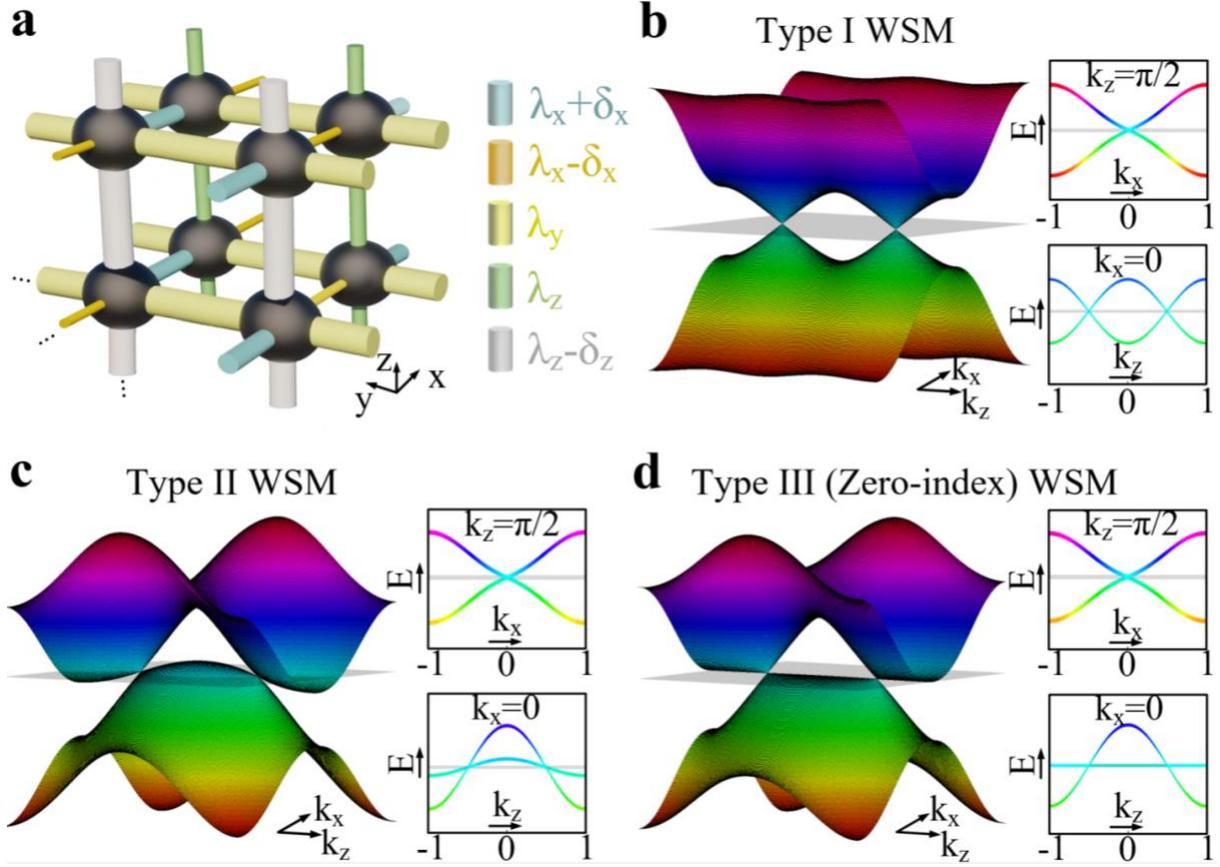

**Fig. 1: Zero-index (type III) Weyl semimetals. a,** We consider a tight-binding model consisting of a set of resonators coupled to each other with specified hopping strengths. **b,** Band structure of the crystal for $\lambda_x=1$, $\lambda_y=2$, $\lambda_z=2$, $\delta_x = 0.5$ and $\delta_z = 0$, calculated at the plane $k_y = 2\pi/3$. The band structure supports a pair of gap-closing Weyl points (WPs). The conical dispersion around the WP is not tilted, leading to a point-like Fermi surface (type I WSM). **c,** Same as panel b except that the parameter $\delta_z$ is chosen to be $\delta_z = 2\lambda_z$. The energy spectrum at the WP is strongly tilted in this case, so that the Fermi surface transforms from a point into an open surface (type II WSM). **d,** Same as b and c but for $\delta_z = \lambda_z$. In this case, the canonical spectrum around the WP is tilted by the critical angle $\theta_c$, at which the valence and conduction bands start to coexist in energy. The Fermi surface of the corresponding semi-metallic phase is neither a point nor a surface, but is a completely flat line. This gives rise to a zero-index response for the semimetal.

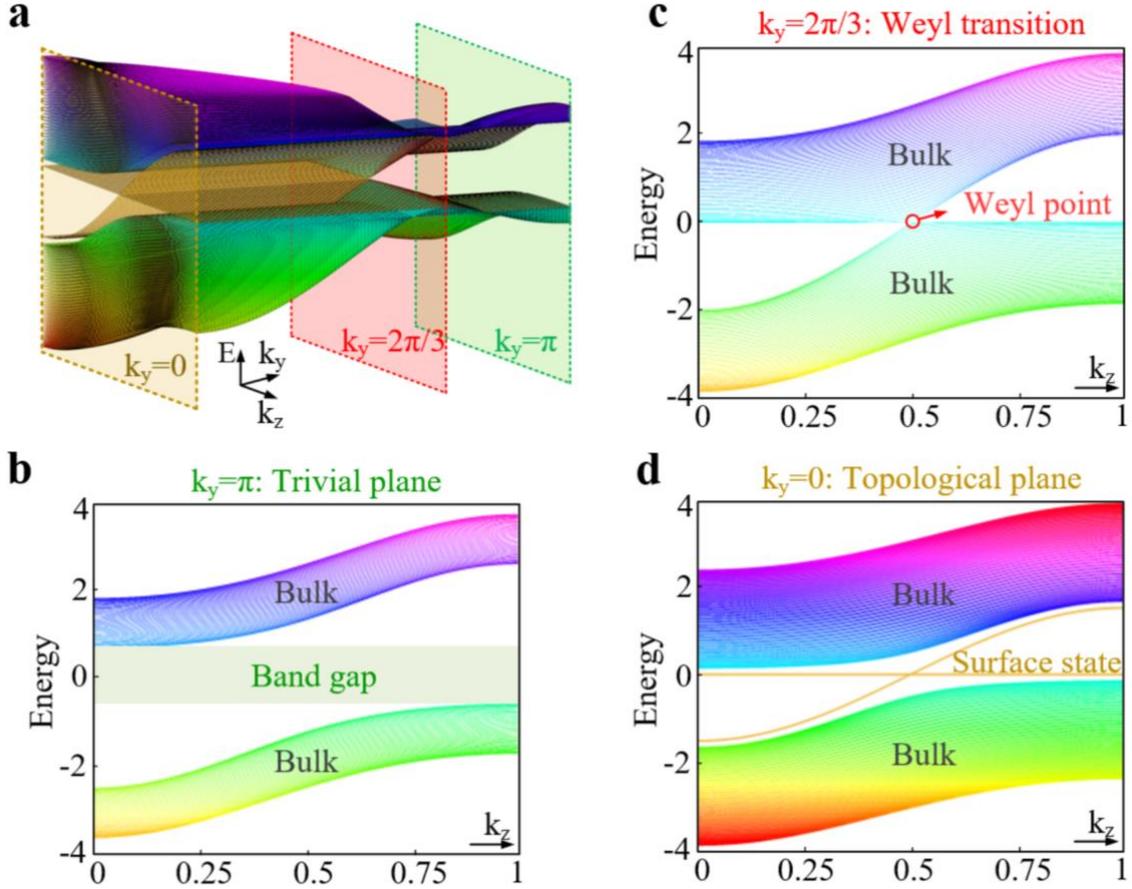

**Fig. 2: Topological boundary states of the proposed zero-index WSM. a,** Dispersion surfaces of the 100×1×1 super-cell of the zero-index WSM. The band structure supports two helical topological boundary states (yellow surfaces). **b,** Band structure of the super cell at $k_y = \pi$, which is completely gapped because the plane section has a zero topological invariant. **c,** Same as panel b but for the plane $k_y = 2\pi/3$. The band structure is gap-less at this plane. **d,** Same as panels b and c except that the band structure is calculated at the plane $k_y = 0$, corresponding to a non-vanishing topological invariant. The band structure exhibits gap-closing topological boundary states. Importantly, one of these boundary states possesses a completely flat dispersion, i.e. a zero group velocity along $z$.

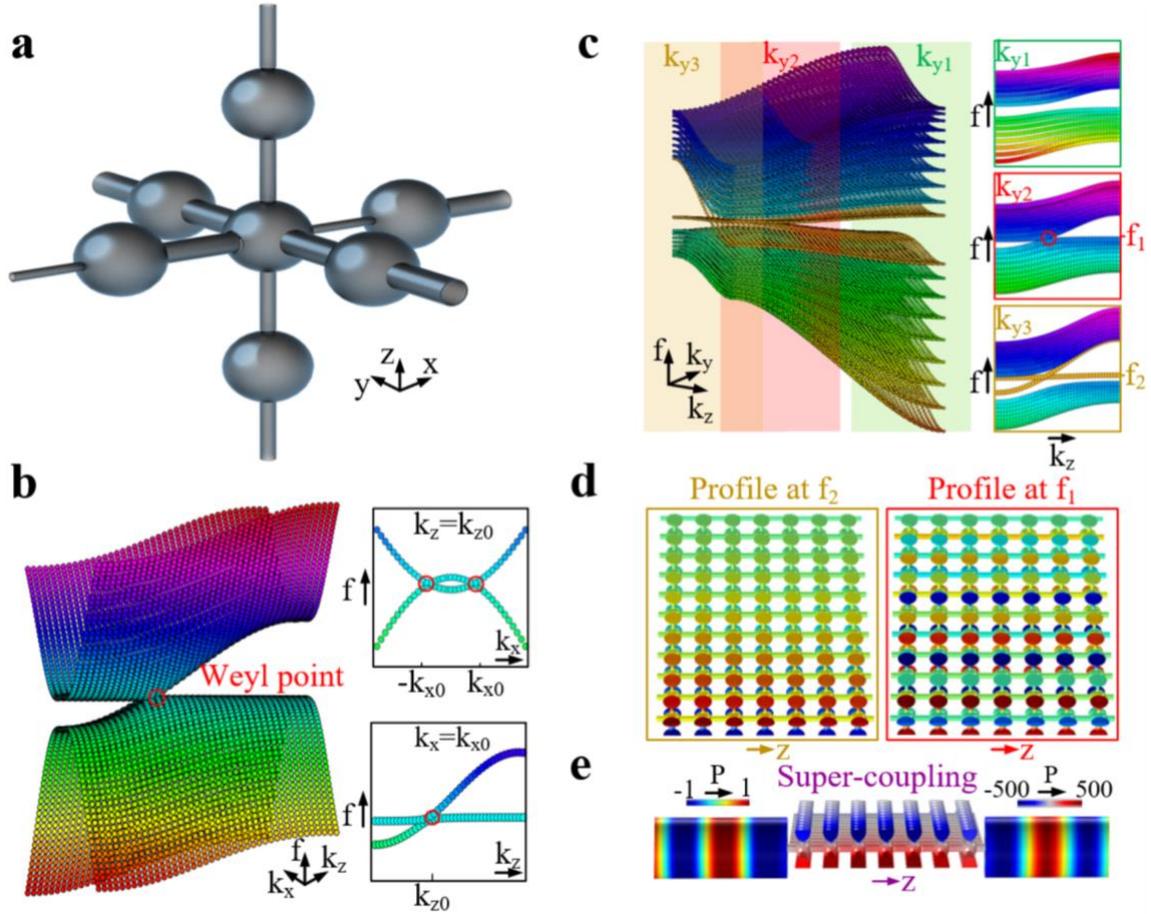

**Fig. 3: Topological acoustic super-coupling induced by zero-index WSMs. a,** We consider a three-dimensional sonic crystal, with Helmholtz resonators connected to each other via acoustic channels with specific widths, realizing a zero-index acoustic WSM. **b,** Band structure of the infinite sonic crystal, resembling the one of the type III WSM described before. **c,** Dispersion surfaces of a supercell of the crystal. Similar to our previous observations, the obtained band structure consists of plane sections with different topological properties. **d,** Left: mode profile of the crystal at the frequency $f_2$. The edge mode has a static-like distribution along the $z$ axis. Right: Pressure distribution at the frequency $f_1$. The bulk mode is also uniform along $z$. **e,** Demonstration of extra-ordinary transmission of sound through the zero-index topological boundary state.

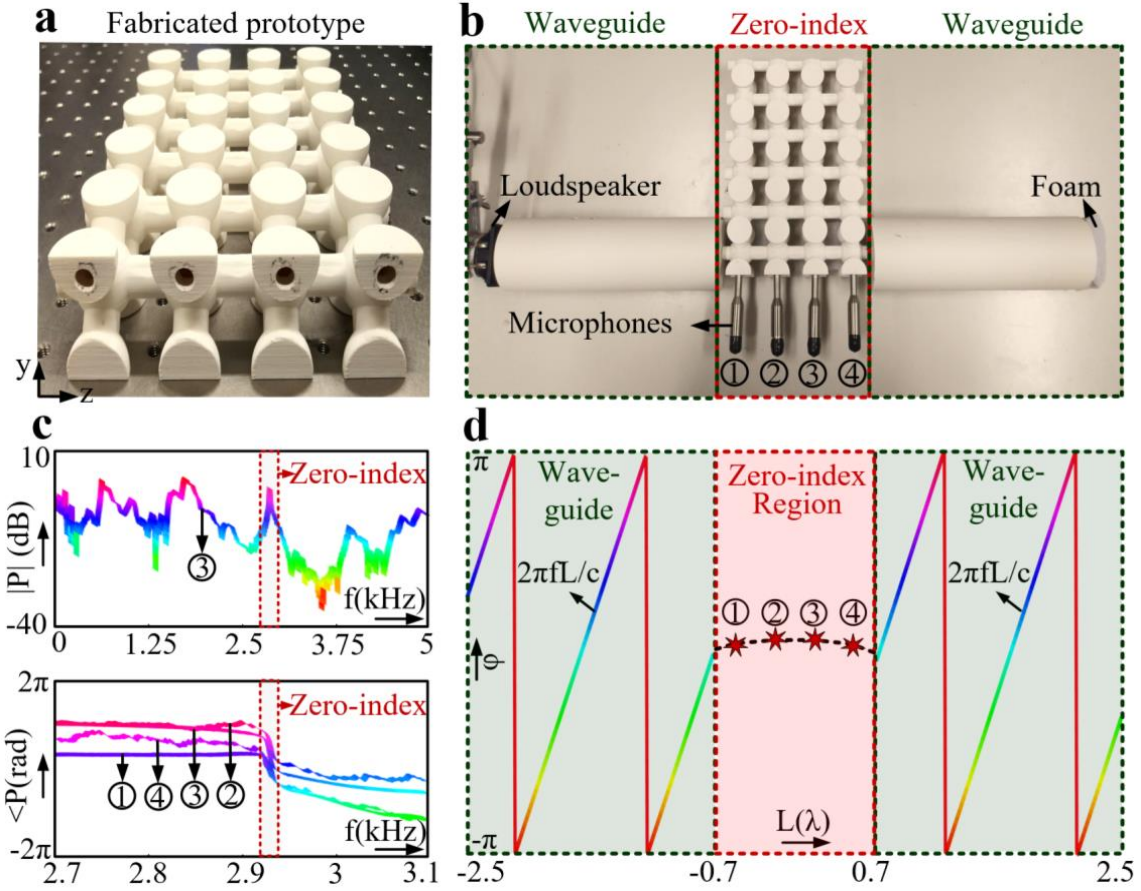

**Fig. 4, Experimental demonstration of anomalous tunneling through the topological zero-index edge mode, a,** Fabricated prototype of a zero-index semimetal, **b,** Experimental setup used to verify the anomalous tunneling through the zero-index topological edge mode. **c,** (Top) Spectrum of the sound pressure level inside the zero-index crystal, measured by the microphone 3. The spectrum exhibits a resonance peak at the frequency of $f_0 = 2920\ Hz$, corresponding to the near-zero index anomalous tunneling. (Bottom) Variation of the pressure phases, measured by each of the four microphones, as a function of frequency. Around $f_0$, the four measured spectra are almost identical, indicating the zero-index character of the crystal within this frequency range. **d,** Variation of the pressure phase along the edge of the crystal (red region), compared to the phase variation in external waveguides ($\varphi = 2\pi f L/c$).

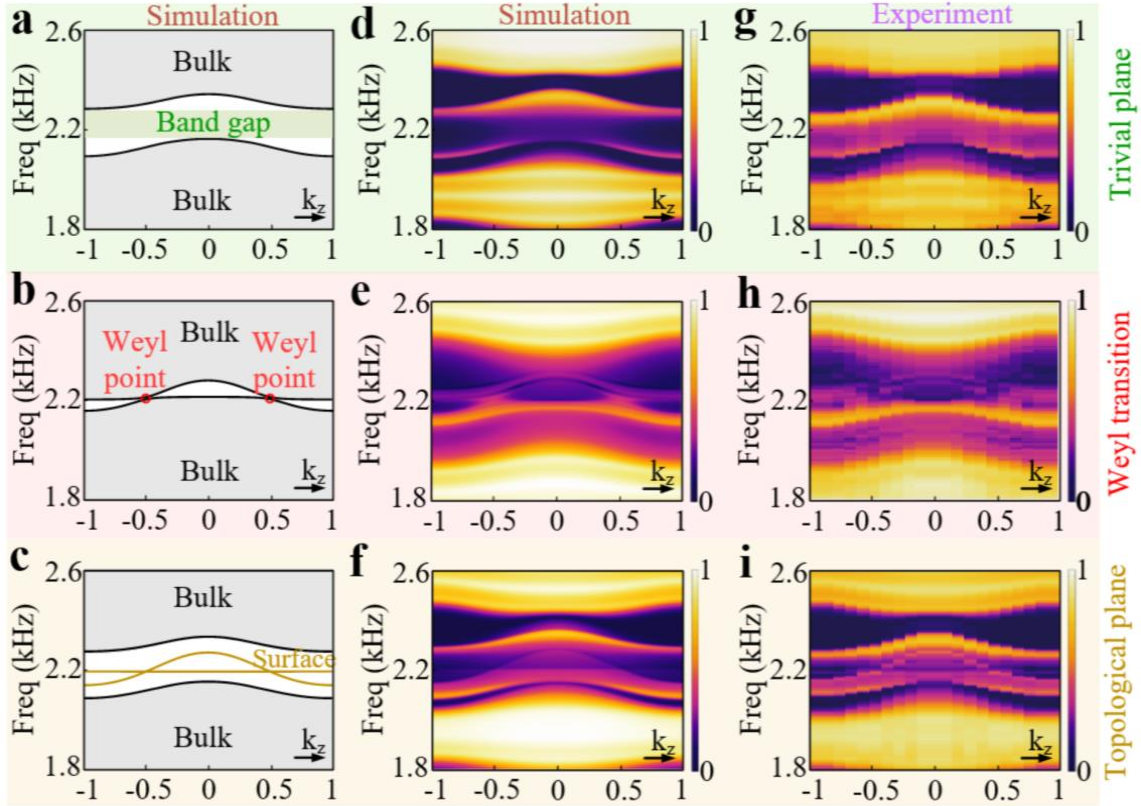

**Fig. 5: Experimental observation of zero-index Weyl semimetals.** We map the tight-binding model represented in Fig. 1 into a one-dimensional phononic crystal [35] with two additional phason degrees of freedom, $\varphi_y$ and $\varphi_z$, taking the role of synthetic Bloch wavenumbers ($k_y$ and $k_z$). **a,** Eigen-frequency spectrum of the sonic crystal at the plane $k_y = 0$, possessing a zero topological index. **b,** Same as panel a but for $k_y = 2\pi/3$ (the plane on transition). **c,** Same as panels a and b except that we set $k_y = \pi$. The band structure exhibits a zero-index topological boundary state. **d,e,f,** Resolution of the obtained eigen-frequency spectra based on the corresponding transmission spectra plotted as a function of frequency and $k_z$, for $k_y = 0$ (panel d), $k_y = 2\pi/3$ (panel e) and $k_y = \pi$ (panel f). **g,h,i,** Corresponding experimental measurements, showing good agreement with simulations.

## Acknowledgments

This work was supported by the Swiss National Science Foundation (SNSF) under Grant No. 172487.